\documentclass{rmf-d}
\usepackage{nopageno,rmfbib,multicol,times,epsf,amsmath,amssymb,cite}
\usepackage[latin1]{inputenc}
\usepackage[]{caption2}
\usepackage{graphicx}
\graphicspath{{figures/}}
\usepackage{url}

\usepackage{placeins}
\usepackage{hyperref}
\usepackage{capt-of}
\newenvironment{Figure}
  {\par\medskip\noindent\minipage{\linewidth}}
  {\endminipage\par\medskip}

\renewcommand{\tabcolsep}{1.35pc}

\def\rmfcornisa{Research OR Education of Physics \hfill\rmf\ {\bf ??} (*?*) ???--???
\hfill MES? A\~NO?}

\def\rmfcintilla{{\it Rev.\ Mex.\ Fis.\/} {\bf ??} (*?*) (????) ???--???}
\clearpage 
\rmfcaptionstyle 
\begin{document}

\title{Systematic treatment of hypernuclear data and application to the hypertriton
\vspace{-6pt}}

\author{
P.~Eckert,$^{a,*}$ 
P.~Achenbach,$^{a,b}$
M.~{Aragon\`es Fontbot\'e},$^a$
T.~Akiyama,$^c$
M.O.~Distler,$^a$ 
A.~Esser,$^a$
J.~Geratz,$^a$ 
M.~Hoek,$^a$
K.~Itabashi,$^c$
M.~Kaneta,$^c$
R.~Kino,$^c$
P.~Klag,$^a$ 
H.~Merkel,$^a$ 
M.~Mizuno,$^c$
J.~M\"uller,$^a$ 
U.~M\"uller,$^a$ 
S.~Nagao,$^c$
S.N.~Nakamura,$^c$ 
Y.R.~Nakamura,$^c$
K.~Okuyama,$^c$
J.~Pochodzalla,$^{a,b}$ 
B.S.~Schlimme,$^a$
C.~Sfienti,$^a$ 
R.~Spreckels,$^a$ 
M.~Steinen,$^b$
K.~Tachibana,$^c$ 
M.~Thiel,$^a$
K.~Uehara,$^c$
and Y.~Toyama$^c$ for the A1 Collaboration
}

\address{$^a$
Institute for Nuclear Physics, Johannes Gutenberg University, Johann-Joachim-Becher-Weg 45, 55128 Mainz, Germany\\
$^*$ email: eckert@uni-mainz.de\\
$^b$ Helmholtz Institute Mainz, GSI Helmholtzzentrum f{\"u}r Schwerionenforschung, Darmstadt,\\ 
Johannes Gutenberg University, 55099 Mainz, Germany\\
$^c$ Graduate School of Science, Tohoku University, Sendai, Miyagi 980-8578, Japan}

\maketitle

\recibido{day month year}{day month year
\vspace{-12pt}}
\begin{abstract}
\vspace{1em} A database is under construction to provide a complete collection of published basic properties of hypernuclei such as $\Lambda$ binding energies, lifetimes, or excitation energies. From these values, averages with related errors are computed in a systematic way. For each property, the overall experimental situation is depicted in form of an ideogram showing the combined probability density function of the measurements. The database is accessible via a dynamic website at \href{https://hypernuclei.kph.uni-mainz.de}{\ttfamily https://hypernuclei.kph.uni-mainz.de} with an user interface offering customizable visualizations, selections, or unit conversions. The capabilities of the database are demonstrated for the puzzling and disputed data situation of the hypertriton.
\vspace{1em}
\end{abstract}

\keys{Hypernuclei; Hypertriton; Binding energies and masses; Lifetimes; Nuclear data compilation
\vspace{-4pt}}
\pacs{ 
    21.10.-k; 
    21.10.Dr; 
    21.10.Tg; 
    21.80.+a; 
    29.87.+g  
\vspace{-4pt}}

\begin{multicols}{2}

\section{Introduction}

Hypernuclear physics is an ongoing field of research that has a rich history of more than 60 years including several eras of diverse experimental approaches to study hypernuclear properties. However, there is no common place where experimental data are collected and combined to averages in such a systematic way as the Particle Data Group (PDG)~\cite{PDG2020} does. This makes the given experimental information hard to access and published averages could be incomplete or outdated.
Therefore, a hypernuclear database was developed to offer a collection of published basic properties of hypernuclei. A key aspect is the combination of measurements in a systematic manner together with a proper treatment of errors and other challenges coming from the multifaceted experimental approaches. At present, the focus is on lifetimes, $\Lambda$ binding energies, and excitation levels. 

For easy data accessibility, the database is placed on a website with an interactive html user interface, where any requests are computed in real-time via JavaScript routines.
The database content is stored in xml files and the whole project is maintained in a git repository. The website is hosted at the Institute for Nuclear Physics of the Johannes Gutenberg University Mainz and can be accessed at \href{https://hypernuclei.kph.uni-mainz.de}{\ttfamily https://hypernuclei.kph.uni-mainz.de} across  multiple browser and device combinations.

\section{Computation of averages}
In the following, the averaging procedures will be explained.

\subsection{Symmetric errors}
%
If a set of data points is reported with purely symmetric errors, the error weighted least-squares procedure is used for the combination to an average. Every mean value $\mu_i$ receives a weight
\begin{equation}
    w_i = \frac{1}{\sigma_i^2}
\label{eq:weight}
\end{equation}
with $\sigma_i$ being the quadratic sum of the statistical error $\sigma_{\mathrm{stat}}$ and the systematic error $\sigma_{\mathrm{syst}}$ of the $i$'th measurement. Normalizing the weights $w_i$ to unity leads to
\begin{equation}
    w_i' = \frac{w_i}{\sum_{j} w_j}.
\label{eq:normweight}
\end{equation}
The average $\Bar{x}$ and its error $\Bar{\sigma}$ are given by the following sums:
\begin{equation}
    \Bar{x} = \sum_{i} w_i' \mu_i
\label{eq:avg}
\end{equation}
\begin{equation}
    \Bar{\sigma} = \sqrt{\sum_i w_i'\,^2 \sigma_i^2}
\end{equation}

\subsection{Asymmetric errors}

In case of data with asymmetric errors $\pm \sigma \to\ ^{+\sigma_+}_{-\sigma_-} $, a procedure 
described by R.~Barlow~\cite{Barlow2004} is used. 
A probability density function {\it pdf}\/ is parameterized via an asymmetric Gaussian distribution
\begin{equation}
    \mathit{pdf}(x) = \frac{1}{\sigma(x)\sqrt{2\pi}} \mathrm{e}^{-\frac{1}{2} \left( \frac{x-\mu}{\sigma(x)}\right)^2}
\label{eq:pdf}
\end{equation}
where $\sigma(x) = \sigma_1 + \sigma_2 (x-\mu)$ is a linear function with $\sigma_1 = 2\sigma_{+} \sigma_{-}/(\sigma_{+} + \sigma_{-})$ and $\sigma_2 = (\sigma_{+} - \sigma_{-})/(\sigma_{+} + \sigma_{-})$ defined such that $\sigma(x-\sigma_{-}) = \sigma_{-} $ and $\sigma(x+\sigma_{+}) = \sigma_{+}$.

Similar to the symmetric case, each measurement receives a weight 
\begin{equation}
    w_i = \frac{\sigma_{1,i}}{(\sigma_{1,i} + \sigma_{2,i}(\Bar{x} - \mu_i))^3}
\end{equation}
and contributes to the average via (\ref{eq:normweight},\,\ref{eq:avg}). However, in this case the weights depend on the average itself, so that it has to be determined in an iterative approach.

The same holds for the errors $\Bar{\sigma}_{+}$ and $\Bar{\sigma}_{-}$ of the average. For their determination, the log-likelihood function 
\begin{equation}
    \ln L = - \frac{1}{2} \sum_i \left(\frac{(x - \mu_i}{\sigma_i (x)}\right)^2 
\end{equation}
is used, with which the errors can be defined by the points where the function is reduced by $\frac{1}{2}$
\begin{equation}
    \ln L(\Bar{x}) - \ln L(\Bar{x} \pm \Bar{\sigma}_{\pm}) = -\frac{1}{2}.
\end{equation}

For the measurements presently in the database, the iterative approximation of the average and its errors was observed to converge quickly, yielding about one order of magnitude in accuracy per iteration, rendering the procedure applicable for on-line processing in a dynamic website. 

It was noticed that the handling of strongly asymmetric errors $(\sigma_+ > 1.5 \sigma_-$ and vice versa) is problematic. In such cases the function $\sigma(x)$ is very steep and hence quickly reaches zero, leading to a pole in \eqref{eq:pdf}. If this pole is too close to the average, the related measurement will receive an oddly high or a negative weight. To prevent this from happening, measurements with such extreme errors are treated with a modified version of the function $\sigma(x)$ which reads
\begin{equation}
    \sigma'(x) =
    \begin{cases}
        \sigma_-,       & \text{if } x\leq \mu - \sigma_-\\
        \sigma_+,       & \text{if } x \geq \mu + \sigma_+.\\
        \sigma(x),      & \text{otherwise}
    \end{cases}
\end{equation}
Here, the parameter $\sigma$ is varied only within the 1-$\sigma$ interval. 

\subsection{Further data treatment}
Several additional procedures were implemented.

\textbf{Error scaling.} Following the PDG~\cite{PDG2020}, the data are handled more conservatively if the $\chi^2$ value
\begin{equation}
    \chi^2 = \sum_i w_i (\mu_i - \Bar{x})^2
\end{equation}
is larger than the number of degrees of freedom ({\it ndf} $=$ number of contributing measurements -- 1). Then a scaling factor $S$ is computed
\begin{equation}
    S = \sqrt{\chi^2/\mathit{ndf}}.
\end{equation}
For $1 < S < 2$, one or more measurements are supposed to have underestimated errors and the error of the average is multiplied by $S$. 
Since this leads to a $\chi^2$ value equal to its statistically expected value $\chi^2 =\mathit{ndf}$, the procedure may be
viewed as being statistically consistent and is also known as Birge ratio algorithm when handling discrepant data~\cite{Taylor1982}.
For $S > 2$, unknown effects are assumed to be in place 
and the average cannot be determined reliably. 

\textbf{Shared systematics.} 
The treatment of shared systematic errors in the case of measurements originating from the same experimental apparatus was adopted from the PDG~\cite{PDG2020}.

\textbf{Exclusion of data.} Measurements which contribute to an average with a weight of less than 2\% are considered as obsolete and therefore excluded. The same holds for weights up to 5\% if the related measurement has an unusually high contribution to the $\chi^2$ value. The rationale of this procedure is, that otherwise these measurements dilute or enlarge the scaling factor.

\textbf{Missing systematic errors.} Earlier binding energy measurements using the emulsion technique have been assigned an error of $\sigma_{\text{syst}} = 40\,\mathrm{keV}$, following an estimation by D. H. Davis~\cite{Davis1992,Davis2005}.

\section{Application to the hypertriton}

The lightest known hypernucleus -- the
bound state of a proton, a neutron and a $\Lambda$ that is called the
hypertriton -- plays a major role in hypernuclear physics and its measured properties are widely used as input for theoretical calculations.

The capabilities of the database are demonstrated in this case, e.g.\ by combining certain data sets, such as from different data taking periods or experimental techniques, and by producing a combined probability density function (\,{\it pdf}\,) from these.  

\subsection{Lifetime  evaluation}
%
In recent relativistic heavy-ion collision experiments, discrepant data of the hypertriton lifetime were extracted,
$142^{+24}_{-21} \pm 29$\,ps 
by the STAR collaboration~\cite{STAR2018}
and $242^{+34}_{-38} \pm 17$\,ps by the ALICE collaboration~\cite{ALICE2019}. Additional conflicting values from this experimental method add to the confusion, see Table~\ref{tab:lifetime} for a full account of the world data set. In earlier imaging measurements a similarly wide spread of values has been measured~\cite{Bohm1970,Keyes1973}, albeit with larger errors. Underestimated systematic errors might have contributed to create this unsatisfactory situation, which has been dubbed the hypertriton lifetime puzzle. Figure~\ref{fig:lifetime} shows the {\it pdf}\/ for the full set of available measurement as well as for selected sub-sets to illustrate that the determination of an average is not very robust. 

\subsection{Binding energy evaluation}

A recent measurement of the hypertriton binding energy has been reported by the STAR Collaboration~\cite{STAR2020}, which is three times larger than the previously known value from emulsion data~\cite{Juric1973,Bohm1968}.\hfill In\hfill contrast,\hfill a\hfill preliminary\hfill value\hfill shown\hfill by\hfill the
\begin{Figure}
  \centering
  \includegraphics[width=.99\textwidth]{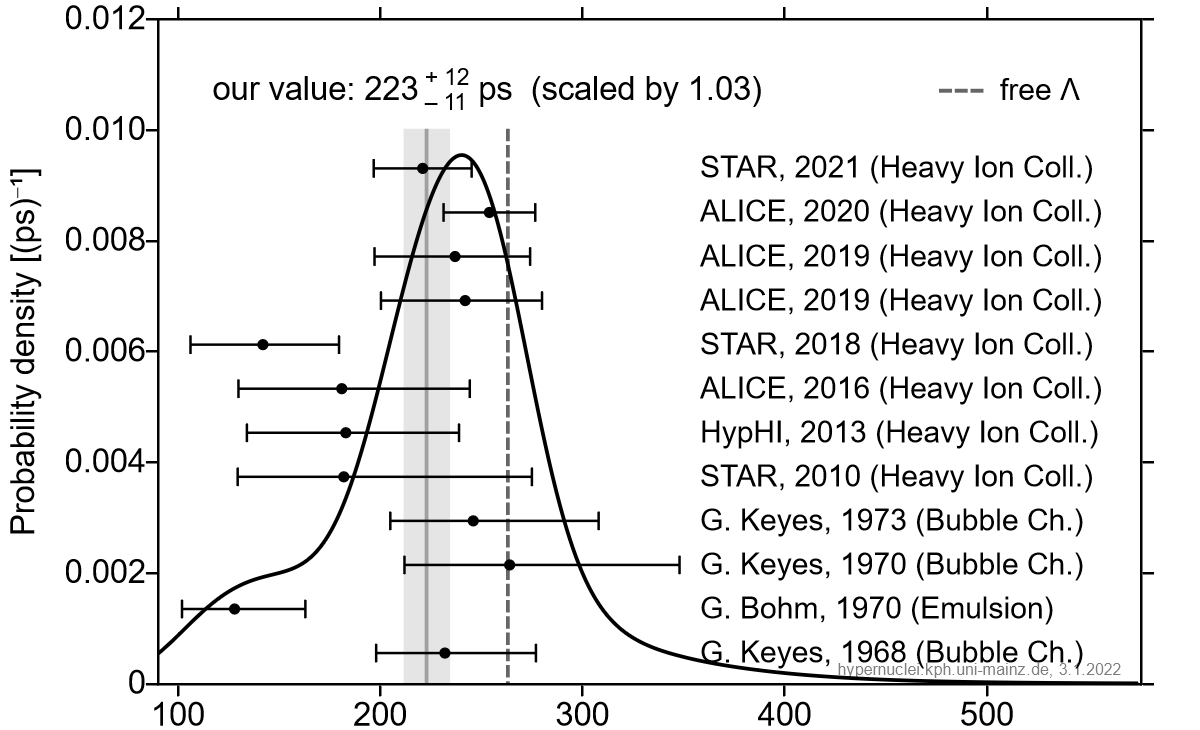}
  \includegraphics[width=.99\textwidth]{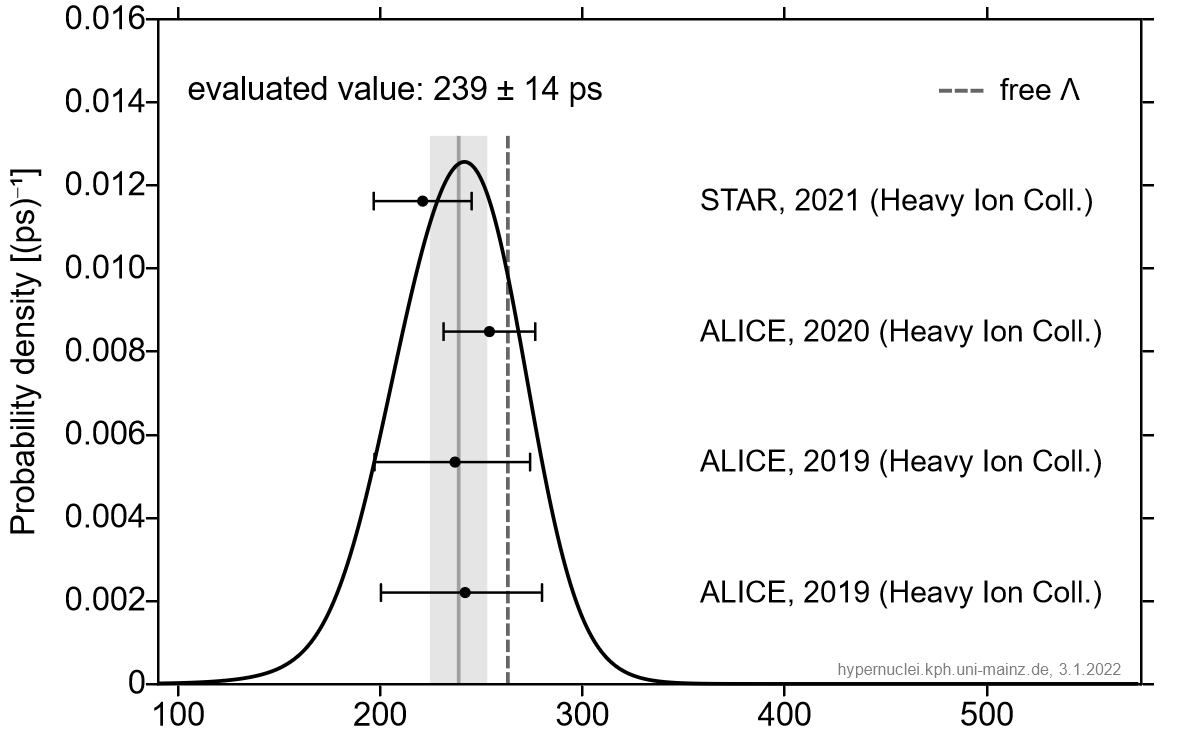}
  \includegraphics[width=.99\textwidth]{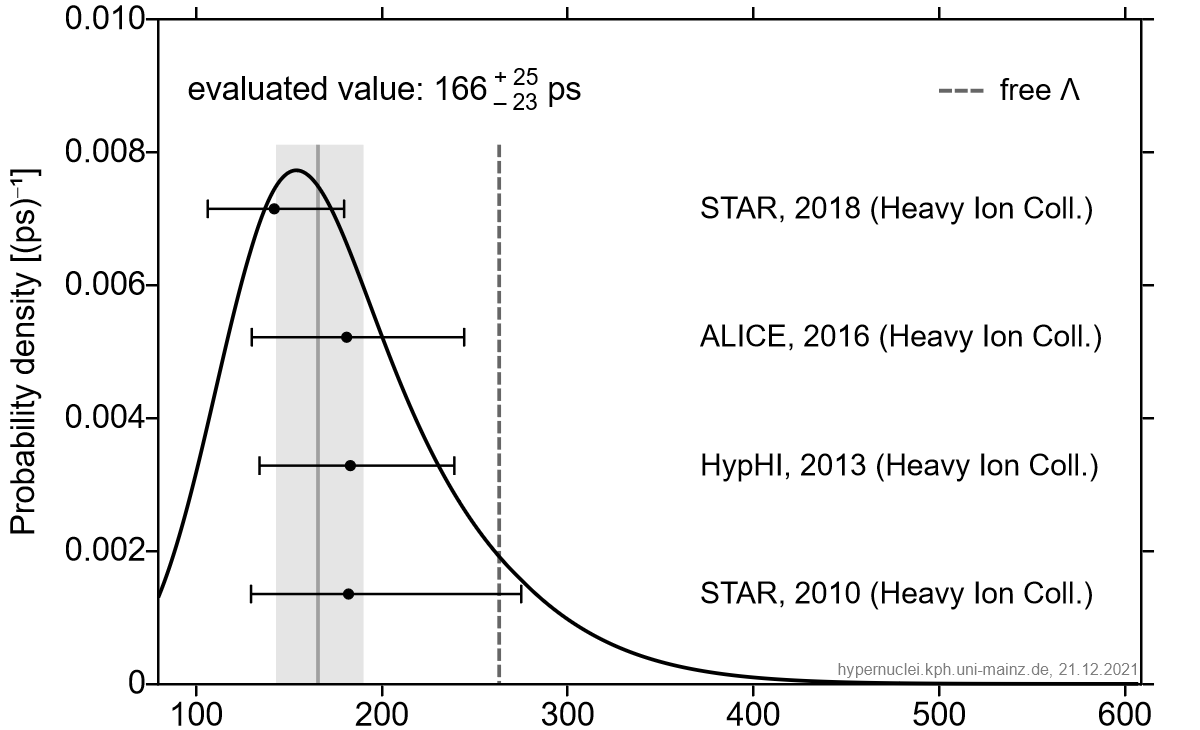}
  \includegraphics[width=.99\textwidth]{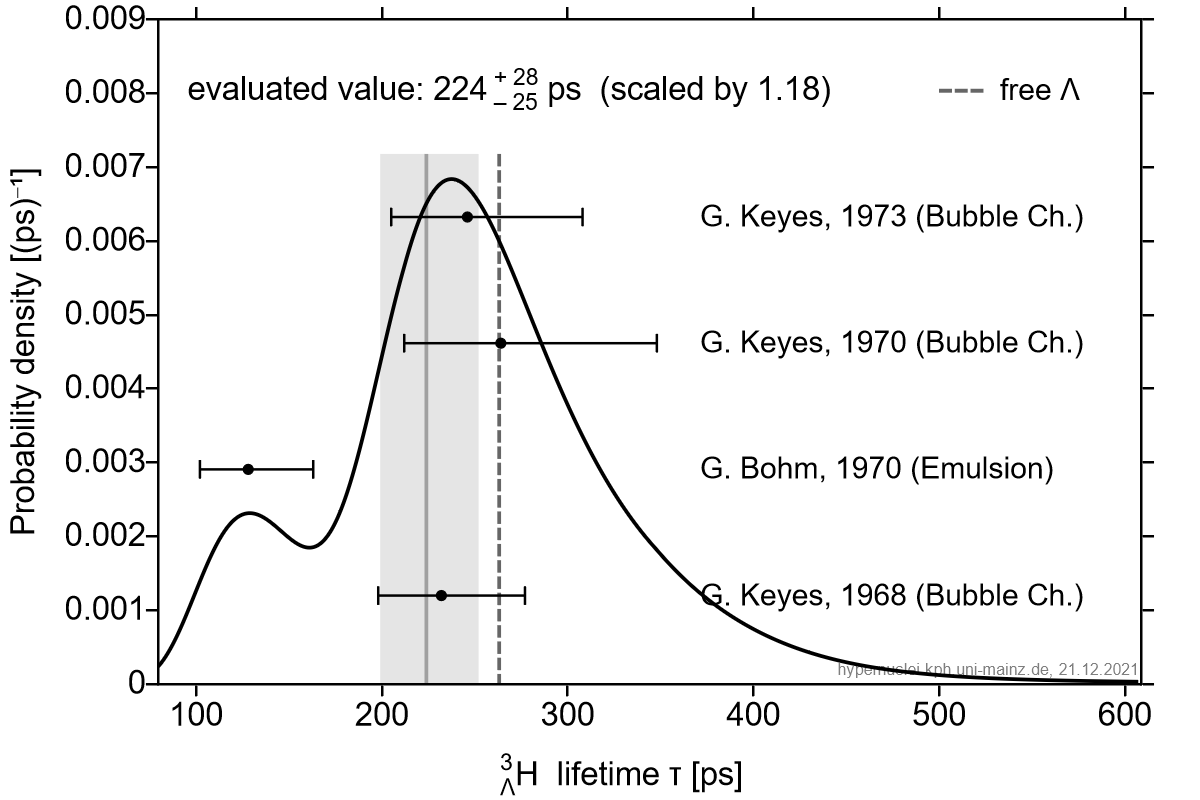}
  \captionof{figure}{Measurements and combined {\it pdf}\/ for the hypertriton lifetime. Different data sets have been selected (from top to bottom), for which the averages were calculated: (1) All: $223^{+12}_{-11}$\,ps; (2) Latest data from heavy ion collision experiments: $239\pm 14$\,ps; (3) Earlier data from heavy ion collision experiments: $166^{+25}_{-23}$\,ps; (4) Data from imaging experiments: $224^{+28}_{-25}$\,ps.}
  \label{fig:lifetime}
\end{Figure}
\noindent ALICE Collaboration confirms a small binding energy value~\cite{ALICE2021}.
It can be inferred that systematic uncertainties play a key role in the measurement of hypertriton properties. 

Table~\ref{tab:bindingenergy} lists the world data set of hypertriton binding energy values and Fig.~\ref{fig:binding} shows the {\it pdf}. The data situation is again unsatisfactory: The two new values from heavy ion collision experiments do not lead to a consistent picture while 
earlier extractions from these experiments did not have sufficient accuracy.
Earlier values from other imaging experiments show a large spread. 

\begin{Figure}
  \centering   \includegraphics[width=.99\linewidth]{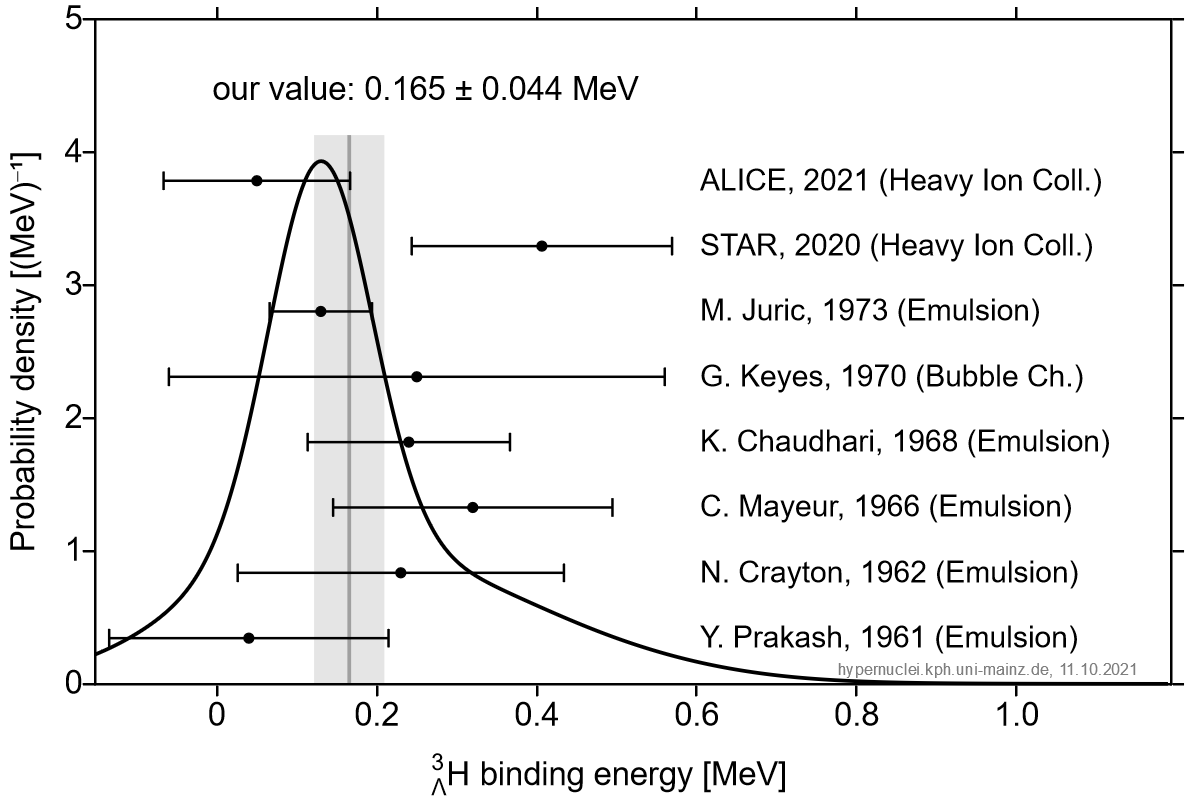}
  \captionof{figure}{Measurements and combined {\it pdf}\/ of the hypertriton $\Lambda$ binding energy.}
  \label{fig:binding}
\end{Figure}

Currently, a new experiment is being prepared at the Mainz Microtron (MAMI) to determine the binding energy of hypertriton
by the spectroscopy of mono-energetic pions from two-body decays of stopped hyperfragments. This technique can yield an unprecedented precision and, combined with a high-luminosity $^7$Li target, achieve a statistical and systematic error of $\sim$ 20\,keV~\cite{Achenbach2018:HADRON2017}. New spectroscopic experiments using $^3$He targets 
are accepted as proposals at Jefferson Lab~\cite{JLabProposal}, J-PARC~\cite{JPARCProposal} and ELPH~\cite{ELPHProposal} which can improve the overall data situation for hypertriton.

\section{Concluding remarks}
%

\begin{Figure}
  \centering
  \includegraphics[width=.7\textwidth]{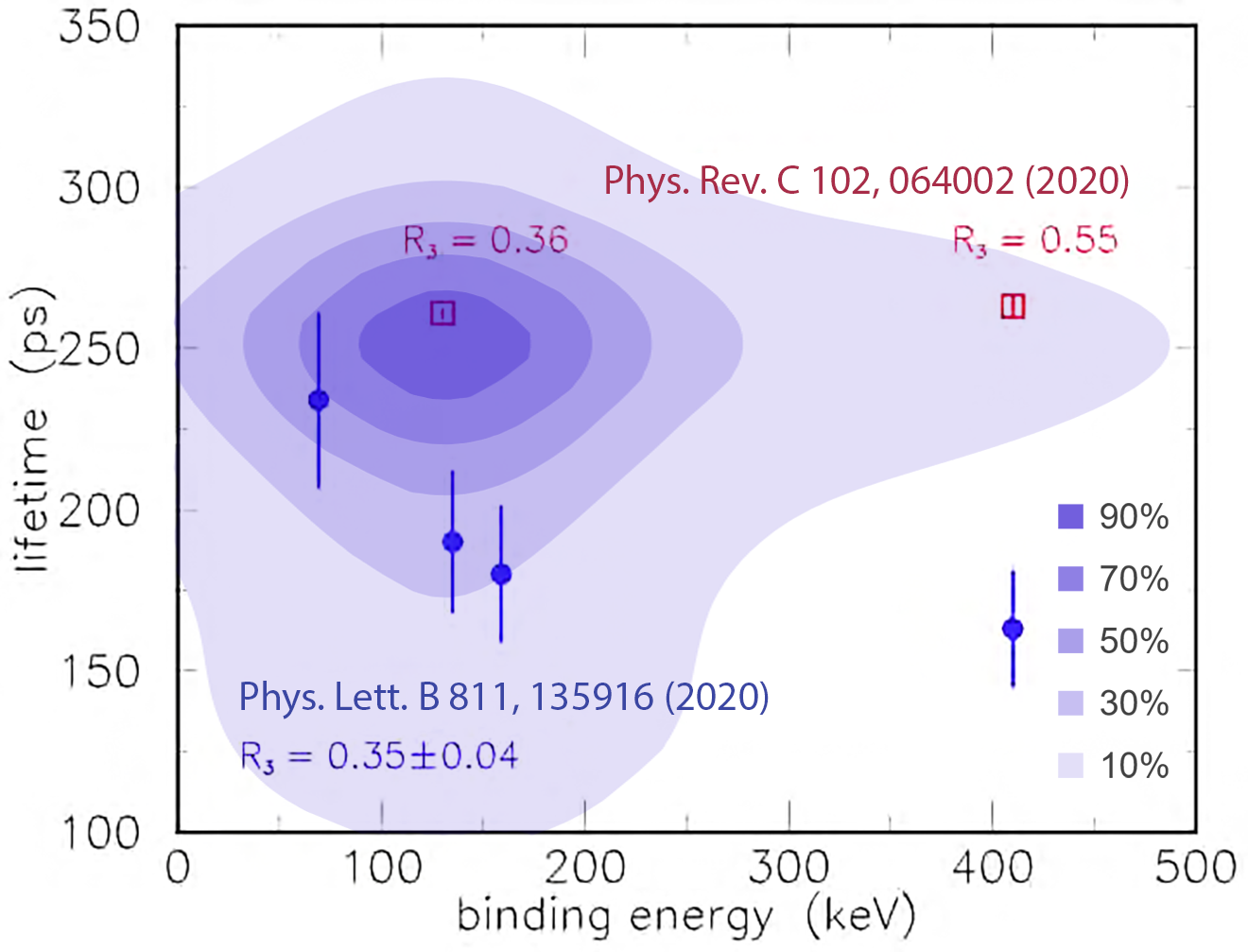}
  \captionof{figure}{Contour lines of the {\it pdf}\/ folding integral for binding energy and lifetime of experimental hypertriton data compared with two recent model predictions~\cite{Hildenbrand2020,PerezObiol2020}.}
  \label{fig:lifetimeAndBinding}
\end{Figure}

Modern microscopic three-body calculations make it possible to connect the hypertriton's binding energy with its lifetime~\cite{Hildenbrand2020,PerezObiol2020}. A relatively small binding implies a small perturbation to the $\Lambda$ wave function inside the hypertriton, and therefore a lifetime close to that of a free $\Lambda$, $263.2 \pm 2.0$\,ps. Fig.~\ref{fig:lifetimeAndBinding} shows the  {\it pdf}\/ folding integral for binding energy and lifetime of the experimental data in comparison with these calculations. 

All figures and tables in this work can be reproduced by using the website. The present data was retrieved on January 3rd, 2022.
It is evident, that the database will be for the benefit of the whole field of experimental and theoretical nuclear strangeness physics. Its usefulness depends in large parts on the interaction between its users and the maintainers. Therefore, data and other valuable information can be send to \texttt{hypernuclei@uni-mainz.de}. More data are constantly reviewed and included in the database. Furthermore, we appreciate comments, criticisms, and suggestions for improvements of any kind.

\vspace{-6.1pt}
\section*{Acknowledgments}
\vspace{-4pt}
This project is supported by the Deutsche Forschungsgemeinschaft, Grant Number PO256/7-1 and the European Union's Horizon 2020 research and innovation programme under grant agreement No.~824093.


\end{multicols}

\begin{table}[htb]
  \longtabletopline\vspace{2pt}
  \caption{World data set of measurements of the hypertriton lifetime in chronological order.}
  \label{tab:lifetime}
  \vspace{2pt}\longtableline
  \begin{center}
    \begin{tabular}{lcllll}
    Reference & Year & Lifetime (ps) & Weight & $\chi^2$ & Technique \\
    \hline
    STAR (prelim.)~\cite{STAR2021} & 2021  & $221 \pm 15 \pm 19$ & 0.21  & 0.01	&  Heavy ion coll. \\
    ALICE (prelim.)~\cite{ALICE2020}& 2020  & $254 \pm 15 \pm 17$ & 0.24  & 1.88	& Heavy ion coll. \\
    ALICE~\cite{Trogolo2019} & 2019  & $237 ^{+33} _{-36}\pm 17$ & 0.08	& 0.13	& Heavy ion coll. \\
    ALICE~\cite{ALICE2019} & 2019	& $242 ^{+34} _{-38}\pm 17$	& 0.07	& 0.22	& Heavy ion coll. \\
    STAR~\cite{STAR2018} & 2018	& $142 ^{+24} _{-21}\pm 29$	& 0.08	& 4.14	& Heavy ion coll. \\
    ALICE~\cite{ALICE2016} & 2016	& $181 ^{+54} _{-39}\pm 33$	& 0.03	& 0.44	& Heavy ion coll. \\
    HypHI~\cite{Rappold2013} & 2013	& $183 ^{+42} _{-32}\pm 37$ & 0.04	& 0.50	& Heavy ion coll. \\
    STAR~\cite{STAR2010} & 2010	& $182 ^{+89} _{-45}\pm 27$ & 0.02	& 0.23	& Heavy ion coll. \\
    DUBNA$^1$~\cite{Avramenko1992} & 1992 & $240^{+170}_{-100}$ & --- & --- & Heavy ion coll. \\
    G. Keyes et al.~\cite{Keyes1973} & 1973	& $246 ^{+62} _{-41}$  & 0.07	& 0.30	& Bubble cham. \\
    G. Keyes et al.~\cite{Keyes1970} & 1970	& $264 ^{+84} _{-52}$ & 0.05 & 0.67	& Bubble cham. \\
    G. Bohm et al.~\cite{Bohm1970}& 1970	& $128 ^{+35} _{-26}$ & 0.04 & 3.19	& Emulsion \\
    R.E. Phillips et al.$^1$~\cite{Phillips1969} & 1969 & $285^{+127}_{-105}$ & --- & --- & Emulsion\\
    G. Keyes et al.~\cite{Keyes1968} & 1968	& $232 ^{+45} _{-34} $ & 0.09	& 0.06	& Bubble cham. \\
    Y.W. Kang et al.$^1$~\cite{Kang1965} & 1965 & $\phantom{1}80^{+190}_{-30}$ & --- & --- & Emulsion \\
    R.J. Prem et al.$^1$~\cite{Prem1964} & 1964 & $\phantom{1}90^{+220}_{-40}$ & --- & --- & Emulsion\\
    L. Fortney$^1$~\cite{Fortney1964} & 1964 & $\phantom{1}63 ^{+50} _{-30} $ & ---	& ---	& Bubble cham. \\
    M.M. Block et al.$^2$~\cite{Block1964} & 1964 &$105 ^{+20} _{-18} $ & ---	& ---	& Bubble cham. \\
    \hline
    Our average & 2022 & $223^{+12}_{-11}$\; ($S = 1.03$) & 1.00 & 11.74 &  ({\it ndf} $=$ 11)
    \end{tabular}
    \longtableline\vspace{1mm}
  \end{center}
  $^1$ excluded due to insignificant weight\\ 
  $^2$ excluded due to missing systematic errors 
\end{table}

\begin{table}[htb]
  \longtabletopline\vspace{2pt}
  \caption{World data set of measurements of the $\Lambda$ binding energy in hypertriton in chronological order. The emulsion data have been assigned an additional systematic error as estimated by D. H. Davis~\cite{Davis1992,Davis2005}.}
  \label{tab:bindingenergy}
  \vspace{2pt}\longtableline
  \begin{center}
    \begin{tabular}{lcllll}
    Reference & Year & Binding energy (keV) & Weight & $\chi^2$ & Technique \\
    \hline
    ALICE (prelim.)~\cite{ALICE2021} & 2021	& $\phantom{4}50 \pm 60 \pm 100$	& 0.14	& 0.98 	& Heavy ion coll.	\\
    STAR~\cite{STAR2020} & 2020	& $406 \pm 120 \pm 110$	& 0.07	& 2.19	& Heavy ion coll. \\
    M. Juri\v{c} et al.~\cite{Juric1973} & 1973	& $130 \pm 50 \pm 40$	& 0.47	& 0.31	& Emulsion	\\
    G. Keyes et al.~\cite{Keyes1970} & 1970	& $250 \pm 310$			& 0.02	& 0.07	& Bubble cham.	\\
    K. Chaudhari et al.~\cite{Chaudhari1968} & 1969	& $240 \pm 120 \pm 40$	& 0.12	& 0.35	& Emulsion \\
    C. Mayeur et al.~\cite{Mayeur1966} & 1966	& $320 \pm 170 \pm 40$	& 0.06	& 0.78	& Emulsion \\
    N. Crayton et al.~\cite{Crayton1962} & 1962	& $230 \pm 200 \pm 40$	& 0.05	& 0.10	& Emulsion \\
    Y. Prakash et al.~\cite{Prakash1961} & 1961	& $\phantom{2}40 \pm 170 \pm 40$& 0.06  & 0.52	& Emulsion \\
    \hline
    Our average & 2022 & $165 \pm 44$ & 1.00 & 5.30 &  ({\it ndf} $=$ 7)
    \end{tabular}
    \longtableline\vspace{1mm}
  \end{center}
  by-products of lifetime measurements from heavy ion collision experiments were excluded due to insignificant weights
\end{table}

\FloatBarrier
\medline
\begin{multicols}{2}

\bibliographystyle{JHEP_PA}
\bibliography{references}

\end{multicols}

\end{document}